\documentclass[sigplan,screen]{acmart}

\usepackage[T1]{tipa}
\usepackage[tt=false, type1=true]{libertine}
\usepackage[libertine]{newtxmath}
\usepackage[shortcuts]{extdash}
\usepackage{seqsplit}

\usepackage{microtype}

\usepackage{tikz,pgfplots}
\usetikzlibrary{arrows.meta}

\pgfplotsset{every axis/.append style={font=\sffamily\scriptsize}}
\usepackage{float}
\usepackage{tipa}
\usepackage{graphicx}
\usepackage{balance}
\usepackage{stfloats}
\usepackage[caption=false,font=footnotesize,labelfont=sf,textfont=sf]{subfig}
\usepackage{booktabs}
\usepackage[symbol]{footmisc}
\usepackage{makecell}
\usepackage{mdframed}
\usepackage{listings}
\usepackage{bm}
\usepackage{enumitem}
\usepackage{multirow}
\usepackage{nicefrac}
\usepackage{amsmath}

\usepackage{amssymb}
\usepackage{booktabs}
\usepackage{todonotes}
\usepackage{arydshln}
\usepackage{booktabs,xcolor,siunitx}

\usepackage{xcolor}
\usepackage{tcolorbox}
\definecolor{lightgrey}{HTML}{D3D3D3}
\definecolor{commentblue}{HTML}{155196}
\definecolor{emphgreen}{HTML}{4F6900}
\definecolor{emphpink}{HTML}{b5283c}

\newcommand{\myslash}{/\hspace{0pt}}

\def\BibTeX{{\rm B\kern-.05em{\sc i\kern-.025em b}\kern-.08emT\kern-.1667em\lower.7ex\hbox{E}\kern-.125emX}}

\newcommand{\ballnumber}[1]{\tikz[baseline=(myanchor.base)] \node[circle,fill=.,inner sep=1pt] (myanchor) {\color{-.}\bfseries #1};}

\setcopyright{acmcopyright}
\acmPrice{15.00}
\acmDOI{10.1145/3578360.3580253}
\acmYear{2023}
\copyrightyear{2023}
\acmSubmissionID{cc23main-p5-p}
\acmISBN{979-8-4007-0088-0/23/02}
\acmConference[CC '23]{Proceedings of the 32nd ACM SIGPLAN International Conference on Compiler Construction}{February 25--26, 2023}{Montréal, QC, Canada}
\acmBooktitle{Proceedings of the 32nd ACM SIGPLAN International Conference on Compiler Construction (CC '23), February 25--26, 2023, Montréal, QC, Canada}
\received{2022-11-10}
\received[accepted]{2022-12-19}

\emergencystretch=\maxdimen
\begin{document}

\title[A Symbolic Emulator for Shuffle Synthesis on the NVIDIA PTX Code]{A Symbolic Emulator for Shuffle Synthesis\\ on the NVIDIA PTX Code}

\author{Kazuaki Matsumura}
\affiliation{\hspace*{-7pt}\mbox{Barcelona Supercomputing Center (BSC)}}
\email{kmatsumura@nvidia.com}

\author{Simon Garcia De Gonzalo}
\affiliation{\hspace*{-7pt}\mbox{Sandia National Laboratories}}
\email{simgarc@sandia.gov}

\author{Antonio J. Peña}
\affiliation{\hspace*{-7pt}\mbox{Barcelona Supercomputing Center (BSC)}}
\email{antonio.pena@bsc.es}

\renewcommand{\shortauthors}{K. Matsumura, S. G. De Gonzalo, A. J. Peña}

\setlength{\textfloatsep}{10pt}
\setlength{\dbltextfloatsep}{10pt}
\setlength{\abovedisplayskip}{0pt}
\setlength{\belowdisplayskip}{6pt}
\setlength{\abovedisplayshortskip}{0pt}
\setlength{\belowdisplayshortskip}{6pt}

\begin{abstract}
Various kinds of applications take advantage of GPUs through automation tools that attempt to automatically exploit the available performance of the GPU's parallel architecture. Directive\-/based programming models, such as OpenACC, are one such method that easily enables parallel computing by just adhering code annotations to code loops. Such abstract models, however, often prevent programmers from making additional low-level optimizations to take advantage of the advanced architectural features of GPUs because the actual generated computation is hidden from the application developer.

This paper describes and implements a novel flexible optimization technique that operates by inserting a code emulator phase to the tail-end of the compilation pipeline. Our tool emulates the generated code using symbolic analysis by substituting dynamic information and thus allowing for further low-level code optimizations to be applied. We implement our tool to support both CUDA and OpenACC directives as the frontend of the compilation pipeline, thus enabling low-level GPU optimizations for OpenACC that were not previously possible. We demonstrate the capabilities of our tool by automating warp-level shuffle instructions that are difficult to use by even advanced GPU programmers. Lastly, evaluating our tool with a benchmark suite and complex application code, we provide a detailed study to assess the benefits of shuffle instructions across four generations of GPU architectures.

\end{abstract}



\maketitle
\vspace{-0.2cm}
\section{Introduction}\label{sec:introduction}
Effectively utilizing the vast amount of computational performance
available in modern supercomputers remains a challenge to this day.
Hardware, middleware, and parallel algorithms should be carefully
orchestrated so that ideal efficiency may be obtained for solving
large real-world problems in high-performance computing (HPC).
Compiler technologies are developed with highly-automated
program optimizations that use domain-specific knowledge and
target architecture specialization to solve a part of this puzzle.
With the end of Moore's Law~\cite{Moore1998} approaching,
the focus on supercomputing technology
is shifting toward even more specialized accelerators,
which in turn increases their complexity. This trend further
signifies the importance of compiler technology to relieve
programmers from the burden of understanding the complex
architecture of modern accelerators to be able to efficiently
optimize their applications.

Currently, Graphics Processing Units (GPUs) are the most widely
adopted accelerator technology, as these are present in seven out of
the top 10 systems in the TOP500 list~\cite{top500}. GPUs work for
accelerating application execution time through their highly parallelized
yet cooperative architecture. To benefit the most from GPUs, however,
programmers must be proficient in writing complex low-level GPU code, often a
largely time-consuming task. To overcome the complexity of low-level
GPU code development, pragma-based programming models such as
OpenACC/OpenMP~\cite{openacc, openmp} have been
developed or adapted to be able to automatically
retarget existing code for acceleration. Although these automation
tools have improved the utilization of GPU acceleration by many
different types of applications, they lack the ability to benefit from
low-level architecture\-/specific optimizations. One such type of
optimizations is the use of warp-level primitives, which have been
available since NVIDIA Kepler GPUs. Warp-level primitives, such as
shuffle operations, may be used to fill a gap between threads and
thread-blocks working as collaborative mechanisms, instead of relying on
shared and global memory accesses.

The main operation across the warp is the {\em shuffle}, which
delivers computed elements to neighbor threads to suppress the
redundancy of computation and memory accesses. However, as many
existing efforts~\cite{simon2019,swizzle,systolic,register-cache} have
demonstrated, those
primitives often require  non-trivial modification of algorithms in
the fundamental part of their codes. 
Since the latency of the shuffle is similar to that of shared memory
loads~\cite{systolic} (apart from storing and synchronization), 
it may serve as a cache system, holding data in
 registers~\cite{register-cache}. However, the effectiveness of this technique is still unknown when disregarding domain\-/specific knowledge.

Our work provides a middle-end environment to extend the code of the
NVIDIA GPU assembly PTX and enables, for the first time in the
literature, {\em automatic} shuffle synthesis to {\em explore the opportunity} of this operation.
Our environment, {\bf PTXASW}\footnote[4]{The artifact is available at \url{https://github.com/khaki3/ptxas-wrapper}.} (Wrapper of PTX optimizing ASsembler),
addresses the entire computational flow of PTX, leveraging a symbolic emulator that can symbolically extract memory-access patterns.
We introduce a Satisfiability Modulo Theories (SMT) solver to prune avoidable control flows
while tracking down the register update.

Following the emulating results,
PTXASW utilizes the solver and detects the global-memory loads that are possible to be covered by the shuffle operation.
Around those loads, additional instructions are implanted,
while supporting corner cases and circumventing overheads.
We conduct the shuffle synthesis on an OpenACC benchmark suite, a
directive-based programming model having no user exposure to warp-level instructions.
Our implementation functions as a plugin of the compilation tool yielding moderate overhead.

Applying our technique, we find various opportunities to enable the shuffle over the original code of the benchmarks.
The performance improvement achieved is up to 132\% with no user intervention on the NVIDIA Maxwell GPU.
Additionally, based on the results of the experiments using several generations of GPUs,
we analyze the latency caused for the shuffle operations to provide
guidelines for shuffle usage on each GPU architecture. In summary, the
contributions of our work are:

\begin{enumerate}[topsep=0.4em, leftmargin=0.2in,rightmargin=0.05in, itemsep=4pt]
\item We create a symbolic emulator to analyze and optimize GPU
  computing code, equipped with an SMT solver for the comparison of symbolic expressions, induction variable recognition for loops, and various optimizations to reduce overheads.

\item Through symbolic analysis, we {\em automatically} find the
  possible cases to utilize the shuffle operation, which previously
  required in-depth domain knowledge to be applied. Then, we synthesize those to the applications, while avoiding expensive computation.

\item Using a directive-based programming model, we generate various
  shuffle codes on several generations of GPUs and show the cases
  that attain performance improvement {\em with no manual effort}.

\item We show the latency breakdown of the optimization on each GPU
  architecture and provide general guidelines for the use of shuffle operations.
\end{enumerate}

Our work is the first attempt at general utilization of shuffles.
Although manual warp-level operations often contributed to
domain-specific optimizations, the metrics to be addressed by
warp-level efforts have not been studied. Even when computation or
memory accesses are reducible, the trade-offs have remained unknown to
date, especially when thread divergence is involved.

The rest of the paper is structured as follows.
Section~\ref{sec:background} provides the necessary background on GPUs
for general-purpose computing, PTX code, and shuffle operations.
Section~\ref{sec:overview} provides a high-level overview of our work.
Sections~\ref{sec:Symbolic} and~\ref{sec:synthesis} describe our
symbolic emulator and shuffle synthesis, while
Section~\ref{sec:method} details our overall methodology.
Sections~\ref{sec:evaluation} and~\ref{sec:analysis} provide the
results of our experimental evaluation and in-depth analysis.
Section~\ref{sec:related} discusses previous related work and
Section~\ref{sec:conclusion} provides concluding remarks.

\vspace{0.2cm}
\vspace*{-1em}
\section{Background}\label{sec:background}

This section provides the necessary background on GPUs for general-purpose computing,
low-level PTX code, and warp-level shuffle operations.

\subsection{GPUs}

\begin{figure}[b]
\vspace{5pt}
\begin{lstlisting}[caption={Addition kernel in CUDA}, label={lst:cuda}]
__global__ void add(
  float *c, float *a, float *b, int *f) {
    int i = [*threadIdx.x*] + [*blockIdx.x*] * [*blockDim.x*];
    if (f[i]) c[i] = a[i] + b[i];       }
\end{lstlisting}
\begin{lstlisting}[caption={Addition kernel in PTX (simplified)}, label={lst:ptx}]
.visible .entry add(.param .u64 c, .param .u64 a,
                    .param .u64 b, .param .u64 f){
/* Variable Declarations */ .reg .pred %p<2>;
.reg .f32 %f<4>;.reg .b32 %r<6>;.reg .b64 %rd<15>;
/* PTX Statements */
ld.param.u64 %rd1, [c];    ld.param.u64 %rd2, [a];
ld.param.u64 %rd3, [b];    ld.param.u64 %rd4, [f];
cvta.to.global.u64 %rd5, %rd4;
mov.u32 %r2, %ntid.x;       mov.u32 %r3, [*%ctaid.x*];
mov.u32 %r4, [*%tid.x*]; mad.lo.s32 %r1, %r3, %r2,%r4;
mul.wide.s32 %rd6, %r1, 4; add.s64 %rd7,%rd5,%rd6;
// if (!f[i]) goto $LABEL_EXIT;
ld.global.u32 %r5, [%rd7]; setp.eq.s32 %p1,%r5,0;
[*@%p1*] bra [*$LABEL_EXIT*];
// %f3 = a[i] + b[i]
cvta.u64 %rd8, %rd2; add.s64 %rd10, %rd8, %rd6;
cvta.u64 %rd11,%rd3; add.s64 %rd12, %rd11,%rd6;
ld.global.f32 %f1, [%rd12];
ld.global.f32 %f2, [%rd10]; add.f32 %f3, %f2, %f1;
// c[i] = %f3
cvta.u64 %rd13,%rd1; add.s64 %rd14, %rd13,%rd6;
st.global.f32 [%rd14], %f3;
[*$LABEL_EXIT:*] ret;                                }
\end{lstlisting}
\vspace{-8pt}
\end{figure}

A Graphics Processing Unit (GPU), is a massively parallel accelerator architecture having with several computational and communication layers.
The minimum execution unit is a {\it thread}.
Each thread can collaborate with other threads bound to a certain {\it thread-block}
and {\it grid}, through per-block shared memory and\slash or grid-wise
global memory.
The architecture is composed of many {\it streaming multiprocessors
  (SMs)}, which execute distributed thread-blocks in groups of threads
(usually 32), called {\it warps}.
Using inner parallel processing units, the SM takes advantage of instruction-level parallelism (ILP), as well as parallelism among warps and thread-blocks.
Since the memory-access latency increases through the levels of the memory hierarchy, the concept of locality is highly respected for performance,
while locality optimizations bring additional synchronization and
resource use to programs.  %
Warp\-/level primitives, available since the NVIDIA Kepler generation of GPUs, allow for the communication among threads within the same warp~\cite{kepler}, avoiding access to either shared or global memory.

All threads execute the same program code, known as GPU {\it kernels}, customarily written in CUDA~\cite{cuda} for NVIDIA GPUs, in a single-instruction
multiple-data fashion. Threads operate on different data, specified in
kernels by programmers, deriving from thread and thread-block identifiers.
Kernels accept arguments, and the number of threads and thread-blocks
is specified as variables.

\subsection{NVIDIA PTX}

User-level code implemented manually in CUDA or OpenACC is brought to
execution on GPUs through NVIDIA PTX~\cite{ptx}, a virtual machine and
ISA for general-purpose parallel thread execution. PTX programs
feature the syntax and sequential execution flow of assembly language.
Thread-specific variables are replicated to be run over SMs in
parallel using the same program but different parameters. Since the
actual machine code (SASS) cannot be modified from official
tools~\cite{turingas}, PTX is the nearest documented and standard GPU
code layer that may be modified.

PTX code consists of kernel and function declarations.
Those have parameters and instruction statements along with variable declarations, labels, and predicates.
Listing~\ref{lst:ptx} provides the CUDA-generated PTX kernel from Listing~\ref{lst:cuda}.
Variable declarations from several data spaces and types correspond to the usage of on-chip resources, especially registers.
Accepting options and types (e.g. \texttt{.eq}, \texttt{.s32}),
PTX instructions leverage defined registers and compute results,
while some of these enable access to other resources (e.g., \texttt{ld.global.u32}).
Predicates ({\bfseries\color{emphpink}\texttt{@\%p1}}) limit the execution of the instructions
stated under them, which may lead to branching based on the thread-specific values, such as thread and thread-block IDs ({\bfseries\color{emphpink}\texttt{\%tid.x}}, {\bfseries\color{emphpink}\texttt{\%ctaid.x}}).
Labels (e.g., {\bfseries\color{emphpink}\texttt{\$LABEL\_EXIT}}) are branch targets
and allow backward jumps that may create loops.

\subsection{Shuffle Operation}\label{subsec:shuffle}

In GPU architectures prior to NVIDIA Kepler, each sequential execution
of a given thread was allowed to transfer data to another thread only
through non\-/local memories, accompanied by a block\-/level or
grid\-/level synchronization barrier. Modern GPU architectures now
support additional data sharing within warps. Intra-warp communication
is performed via shuffle operations. Listing~\ref{lst:shuffle} shows
the \texttt{{\bfseries\color{emphgreen}shfl}.sync} instruction in PTX,
in which data gets shifted unidirectionally (\texttt{.up},
\texttt{.down}) across the threads of the warp, swapped in a butterfly way (\texttt{.bfly}), or exchanged by precise indexing (\texttt{.idx}).

In the unidirectional shuffle, the delta part, which has no source lane from the same warp, will be unchanged and obtain a false value in the resultant predicate ({\bfseries\color{emphpink}\texttt{\%p1}});
only the active threads (\texttt{\%mask}) of the same control flow
participate in the same shuffle. Inactive threads or threads from
divergent flows produce neither valid results nor predicates to destination lanes.
Each operation is accompanied by the warp-level synchronization, some of which are optimized away during compilation.
While shuffle instructions allow for sub-warp granularity,
our paper focuses on the unidirectional instruction with 32 threads using 32-bit data,
as applying sub-warp granularity to applications tends to feature corner cases and suffers from exception handling for intricate patterns.

\begin{figure}[h]
\begin{lstlisting}[caption={The use of \texttt{shfl.sync} in PTX}, label={lst:shuffle}]
activemask.b32 %mask;
// val[warp_id] = %src; %dst = val[warp_id-%i]
shfl.sync.up.b32   %dst1|%p1, %src, %i,  0, %mask;
// val[warp_id] = %src; %dst = val[warp_id+%i]
shfl.sync.down.b32 %dst2|%p2, %src, %i, 31, %mask;
// val[warp_id] = %src; %dst = val[warp_id^%i]
shfl.sync.bfly.b32 %dst3|%p3, %src, %i, 31, %mask;
// val[warp_id] = %src; %dst = val[%i]
shfl.sync.idx.b32  %dst4|%p4, %src, %i, 31, %mask;
\end{lstlisting}
\vspace{-10pt}
\end{figure}

Table~\ref{tab:latencies} shows the latencies (clock cycles) of shared
memory (SM; no-conflict) and L1 cache as reported by~\cite{dissecting},
besides that of shuffle, from a microbenchmark based
on~\cite{cudabmk}. In the table,
Kepler is NVIDIA Tesla K80, Maxwell is M60, Pascal is P100 and
Volta is V100, while Tesla K40c/TITAN X are used for the shuffle of
Kepler/Maxwell. This table reveals that shuffle brings benefits over
shared memory as a communication mechanism when data movement is not
redundantly performed, so storing and synchronization are
avoidable. In particular, latencies of L1 cache on Maxwell/Pascal are higher compared to Kepler/Volta, which integrate shared memory with L1 cache. Those allow the shuffle to be utilized as a register cache for performance improvement, but the engineering efforts in order to modify the fundamental parts of parallel computation are considerably high.

\begin{table}[h]
\begin{center}
\vspace*{-0.25cm}
\begin{tabular}{c|ccc}
name & Shuffle (up) & SM Read & L1 Hit \\\hline
\textbf{Kepler} & 24 & 26 & 35 \\
\textbf{Maxwell} & 33 & 23 & 82 \\
\textbf{Pascal} & 33 & 24 & 82 \\
\textbf{Volta} & 22 & 19 & 28 \\
\end{tabular}
\end{center}
\caption{\label{tab:latencies}Latencies (clock cycles) as reported by~\cite{dissecting,cudabmk}}
\vspace{-0.5cm}
\end{table}

\vspace{-0.2cm}

\begin{figure*}[t]
  \centering
  \includegraphics[width=0.99\textwidth]{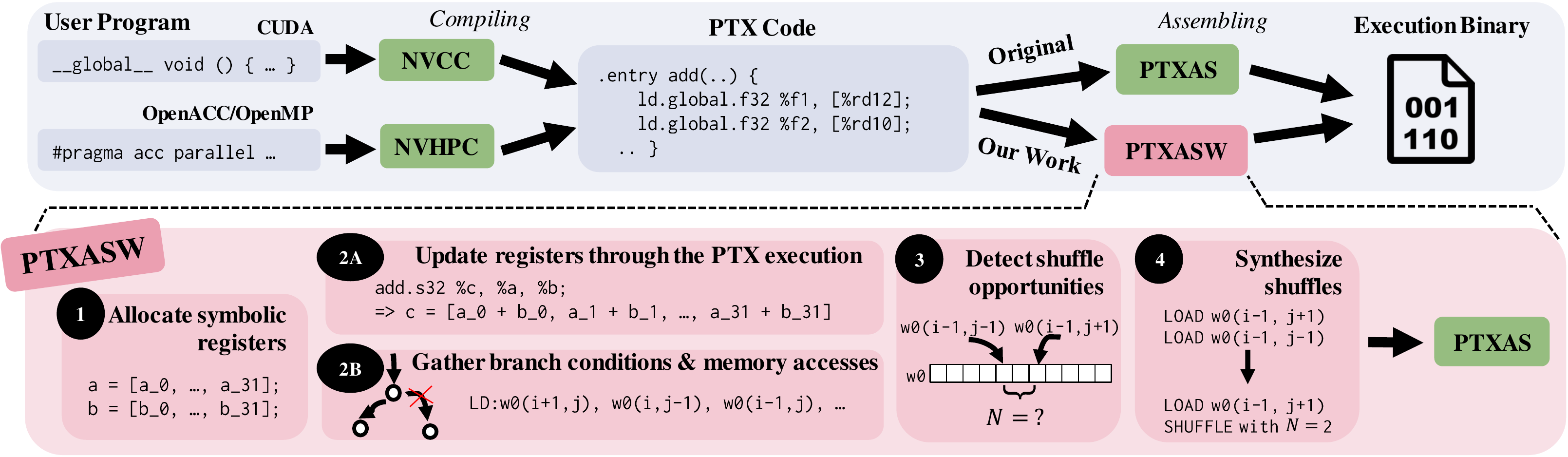}
  \vspace*{-0.2cm}
  \caption{Overview of PTXASW}
  \label{fig:overview}
  \vspace*{-0.1cm}
\end{figure*}

\section{Overview}\label{sec:overview}

Our work PTXASW can substitute the original PTX assembler, which accepts input code from arbitrary sources.
We do not rely on specific information of any certain language or any
certain generation of GPU architecture. %

Figure~\ref{fig:overview} provides a high-level overview of PTXASW's execution flow.
PTXASW primarily aims at shuffle synthesis on PTX code. The input is
produced by user\-/level code compilers, while directive-based
programming models (OpenACC\slash OpenMP) do not expose control over
warp-level operations, and CUDA prevents code extension due to its
code complexities.  %
Once PTXASW inserts shuffles, the resultant code is assembled to GPU binary by the original PTX assembler.

PTXASW emulates the PTX execution based on the input.
Since runtime information is not provided, we employ symbolic evaluation for each operation.
First, \ballnumber{1} register declarations are processed to be mapped in a symbolic register environment (described in Section~\ref{sec:inst}).
Second, \ballnumber{\footnotesize 2A} for each statement of PTX instructions, a corresponding operation is performed to update registers (Section~\ref{sec:inst}). While continuing the execution, \ballnumber{\footnotesize 2B} PTXASW gathers branch conditions for avoiding unrealizable paths (Section~\ref{sec:branch}) and creates memory traces (Section~\ref{sec:memory}).
When the entire emulation is finished, \ballnumber{3}
we discover shuffle opportunities from memory traces (Section~\ref{sec:detect}).
Finally, \ballnumber{4} we insert shuffle operations to the input code (Section~\ref{subsec:codegen}); then, the generated code is consumed by the original PTX assembler.

\vspace{-0.2cm}
\section{Symbolic Emulator}\label{sec:Symbolic}

Analysis of high-level code has posed questions about its applicability to abstract program structures or other user-level languages.
While high-level code analysis may process intact code information, enormous engineering efforts are required just for specific forms within one language~\cite{simon2019, pet}.
Therefore, virtual machines are utilized for providing a cushion
between real architectures and user codes. In particular, analysis and optimization of the virtual-machine code tend to be reusable without the restriction of input types~\cite{polly, legacy, ptx-opt}.

Our work uses PTX as the virtual machine layer and performs general analysis through code emulation.
We introduce symbolic emulation to encapsulate the runtime information in symbol expressions and compute concolic (concrete + symbolic) values for each register.
Although a number of previous work have been conducted on symbolic emulation for the purpose of software testing~\cite{survey},
our work (PTXASW) especially aims at code optimization of memory access on GPUs,
since it is often regarded as one of the bottlenecks of GPU
computing~\cite{systolic}. Those computed values are utilized for code
generation as described in Section~\ref{sec:synthesis}.

\subsection{Instruction Encoding}\label{sec:inst}

Since the subsequent PTX assembler, while generating SASS code, will
eliminate redundant operations and resources, we may
abundantly use registers while not causing register pressure by
unnecessary data movement outside of the static single assignment form
(SSA).
First, PTXASW recognizes variable declarations and prepares a symbolic bitvector of the corresponding size for each register.
Since arithmetic calculation and bitwise operations are supported on
the combination of concrete and symbolic bitvectors, we encode each
PTX instruction as the computation over vectors. For example, addition
for 16-bit vectors is encoded as in the following pseudocode:
\begin{lstlisting}
a = [a_0, a_1, .., a_15]; //a_N is a 1-bit element
b = [b_0, b_1, .., b_15];
c = a + b
  = [a_0 + b_0, a_1 + b_1, .., a_15 + b_15];
\end{lstlisting}

With the \texttt{add} instruction corresponding to the above calculation, we detect the instruction type and source registers (\texttt{\%a}, \texttt{\%b}) and compute the result:
\begin{lstlisting}
add.u16 %c, %a, %b; // dst: %c; src: %a, %b
\end{lstlisting}

Then, having the binding with the name of the destination register (\texttt{\%c}),
we keep the computed value in the register environment.
PTXASW defines each instruction to update the destination registers according to the instruction options and types,
and those registers may be fully concrete with the movement or computation from constant values.
Also, to support floating-point instructions,
we insert the conversion by uninterpreted functions at loading and storing
bitvectors to and from floating-point data.
Regarding casting operands among integer types and binary types,
truncating or extending is performed based on the PTX specification.
The computational instructions under predicates issue conditional values in registers.
Since registers are not used before initialization, these always have evaluated values, except for special registers, such as thread IDs and uninterpreted functions of loops and memory loads, which are described in following sections.

\begin{figure}[b]
\begin{lstlisting}[caption={Jacobi kernel in Fortran and OpenACC}, label={lst:jacobi}]
_*!$acc  kernels*_ loop independent gang(65535)
_*!$acc*_& present(w0(1:nx,1:ny), w1(1:nx,1:ny)) 
[*do*] j = 2, ny-1
  _*!$acc loop*_ independent vector(512)       
  [*do*] i = 2, nx-1
    w1(i,j)=c0*w0(i,j) + c1*(w0(i-1,j)+w0(i,j-1)+&
      w0(i+1,j)+w0(i,j+1)) + c2*(w0(i-1,j-1)+&
      w0(i-1,j+1)+w0(i+1,j-1)+w0(i+1,j+1))
[*enddo*] [*enddo*]
\end{lstlisting}
\vspace*{-12pt}
\end{figure}

\begin{figure*}[b]
\vspace{7pt}
\begin{lstlisting}[caption={Global-memory trace of Jacobi kernel
    through the symbolic emulation in order. Sign extensions are
    omitted. Numerical numbers, shown in hexadecimal, are originally
    in
    bitvectors. \texttt{\bfseries\color{emphgreen}load}/\texttt{\bfseries\color{emphgreen}loop} are uninterpreted functions for parameter loads having addresses and loop iterators having unique identities, respectively  %
)}, label={lst:load}, emph={[1]load, loop}, emph={[2]tid, ctaid}, emphstyle={[2]\color{emphpink}\bfseries}]
LD: 0xc + (load(param2) + ((((0x1 + %ctaid.x) * load(param6) // w0(i-1, j+1)
                 + ((%tid.x + %ctaid.y << 0x9) + (- load(param5)))) + loop(0, 14)) + loop(0, 53)) << 0x2)
LD: 0xc + (load(param2) + (((load(param6) * (0x3 + %ctaid.x) // w0(i+1, j+1)
                 + ((%tid.x + %ctaid.y << 0x9) + (- load(param5)))) + loop(0, 13)) + loop(0, 52)) << 0x2)
LD: 0x4 + (load(param2) + ((((0x1 + %ctaid.x) * load(param6) // w0(i-1, j-1)
                 + ((%tid.x + %ctaid.y << 0x9) + (- load(param5)))) + loop(0, 14)) + loop(0, 53)) << 0x2)
/* LD: w0(i+1, j-1), w0(i  , j+1), w0(i+1, j  ), w0(i  , j-1), w0(i-1, j  ), w0(i  , j  ) */
ST: 0x8 + (load(param3) + (((%tid.x + %ctaid.y << 0x9)       // w1(i  , j  )
           + loop(0, 57)) + ((- load(param5)) + load(param6) * ((0x2 + %ctaid.x) + loop(0, 21)))) << 0x2)
\end{lstlisting}
\vspace{-5pt}
\end{figure*}

\subsection{Execution Branching}\label{sec:branch}
Branching is caused by jumping to labels under binary predicates that are computed by preceding instructions.
Since inputs and several parameters are unknown at compilation time, unsolvable values of predicates are often observed leading to
undetermined execution flows where computation is boundless.
Thus, we abstract the repeated instructions in the same execution flow.
At the entry point to the iterative code block, we modify each iterator of the block to have uninterpreted functions with unique identities and perform operations only once upon those uninterpreted functions.
Since those uninterpreted functions produce incomparable values, we clip the initial values out and add them to registers containing uninterpreted functions at the block entry, for better accuracy in the case of incremental iterators to be found by induction variable recognition~\cite{giv1, giv2}.

We continue each branching while duplicating the register environment
for succeeding flows. All the flows finish at re-entry to iterative
blocks or at the end of instructions, completing their own results.
The symbolic expressions in predicates used at the prior divergence are recorded as assumptions
while updating those predicates, to have constant booleans in the
register environment, based on whether it is assumed as true.  %
Conflicting values in assumptions are removed according to an SMT solver (Z3~\cite{z3}) when new expressions are added.
If the destination of a new branch can be determined providing assumptions to the solver, unrealizable paths are pruned for faster emulation.
Also, we skip redundant code-block entry bringing the same register environment as other execution flows by memoization, to force new results at each entry.

\subsection{Memory Analysis}\label{sec:memory}

We collect memory loads forwardly through the emulation and express
them by uninterpreted functions accepting addresses and returning data
of corresponding sizes.  %
The trace of memory loads is intervened by memory stores, and both loads and
assumptions are invalidated by stores that possibly overwrite them, using the same mechanism for conflicting assumptions mentioned in Section~\ref{sec:branch}.

Listing~\ref{lst:jacobi} shows a Jacobian kernel implemented in Fortran for GPUs using OpenACC.
Its memory trace is obtained as in Listing~\ref{lst:load}
by PTXASW emulating the PTX code generated by NVHPC compiler 22.3.
The address of each load is symbolically calculated as register values, thus containing uninterpreted functions and special registers.
In the case of divergence, branched flows maintain such traces while sharing the common parts of the original flow.

\vspace{-0.2cm}
\section{Shuffle Synthesis}\label{sec:synthesis}

Mapping programs over thread-level parallelism, while pursuing the
performance of modern complex architectures and ensuring correctness,
is a far--from--easy task.
Most likely, existing GPU programs are already optimized in terms of resource use and scheduling, which does not smoothly allow for further optimization, especially at the low-level code.
The shuffle operation performs at its best when the communication is fully utilized~\cite{swizzle}, but such cases are not common in compiler\-/optimized code or even manually-tuned code in HPC.
The big trouble is corner cases.
Not only halo, but fractional threads emerged from rounding up dynamic input sizes, demand exceptional cases to be operated on GPUs.
While the generality and applicability of GPU shuffle instructions for
all types of applications or computational patterns are yet unknown,
the level of difficulty in manually applying shuffle instructions in
different cases adds further hardness to the already complex task of understanding the true
nature of the performance of shuffle operations.

Hence, we implement automatic shuffle synthesis through PTXASW
to drive the lower-latency operations seen in Section~\ref{subsec:shuffle}, while supporting corner cases
and covering global-memory loads with warp-level communication.
PTXASW is accordingly extended to seek shuffle candidates among loads, and
embed shuffle instructions into code while alleviating
register pressure.

\subsection{Detection}\label{sec:detect}

Warps are comprised of neighboring threads. We do not consider adjacent threads in
non-leading dimensions, since those tend to generate non-sequential
access patterns.
Upon finding a global-memory load, PTXAS compares its load address to those of previous loads found through the same execution flow and not invalidated by any store. If for all threads in a warp the load is overlapped with existing loads, those instructions are recorded as possible shuffle sources.
To utilize a load with an address represented as $A(\texttt{\%tid.x})$ for another having the address $B(\texttt{\%tid.x})$, there must exist an integer $N$ such that $A(\texttt{\%tid.x}+N)=B(\texttt{\%tid.x})$ and $-31 \leqslant N \leqslant 31$. For example, when $N=0$, the load can be fully utilized in the same thread. When $N=1$, we can adapt the \texttt{shfl.sync.down} instruction to convey existing register values to next threads while issuing the original load for the edge case ($\texttt{\%warp\_id}=31$). In the case of the memory trace in Listing~\ref{lst:load}, the load accesses of \texttt{w0(i-1, j+1)} and \texttt{w0(i-1, j-1)} are uniformly aligned with the close addresses to each other, so we can search the variable $N$, which satisfies the above condition, by supplying $N$ along with those addresses to the solver and find $N=-2$.

We make sure that each shuffle candidate has the same $N$ as a shuffle
delta in all the execution flows. This delta must be constant
regardless of runtime parameters. Since the steps of loop iterators in
PTX code could be any size (e.g. NVHPC Compiler uses the thread-block
size), shuffles are detected only in straight-line flows, whereas live
variable analysis is employed to exclude the case in which source
values possibly reflect a different iteration from the destination.
For faster analysis, we construct control-flow graphs before shuffle detection, while pruning unrelated instructions to memory operations and branches, and at the use of the SMT solver, uninterpreted functions are converted to unique variables.

\vspace*{-1em}
\subsection{Code Generation}\label{subsec:codegen}

\begin{figure}[b]
\vspace{5pt}
\begin{lstlisting}[caption={Shuffle synthesis on Jacobi kernel (Upper is original and lower is synthesized code; variable declarations are omitted and the naming is simplified)}, label={lst:synthesis}, escapechar = !]
ld.global.nc.f32 %f4, [%rd31+12];// w0(i-1, j+1)
/* ... */
ld.global.nc.f32 %f7, [%rd31+4]; // w0(i-1, j-1)
!\tikz[remember picture] \node [] (a) {};!
ld.global.nc.f32 %f4, [%rd31+12];
mov.f32 [*%source*], %f4; /* ... */
mov.u32 %wid, %tid.x; rem.u32 %wid, %wid, 32;
activemask.b32 %m; setp.ne.s32 %incomplete, %m, -1;
setp.lt.u32 %out_of_range, %wid, 2;
or.pred %pred, %incomplete, %out_of_range;
shfl.sync.up.b32 %f7, [*%source*], 2, 0, %mask;
@%pred ld.global.nc.f32 %f7, [%rd31+4];
\end{lstlisting}
\begin{tikzpicture}[remember picture, overlay]
  \node [right = 8.07cm of a] (A) {};
  \node [left = 0cm of a] (B) {};
  \node [below left = 0cm and -.25cm of A, font=\bfseries] {PTXASW};
  \draw[dashed] (B.east) edge (A.east);
\end{tikzpicture}
\vspace*{-15pt}
\end{figure}

Warp divergence may be caused by various reasons, including the dynamic
nature of the program execution, which is inconvenient to optimization, where the uniformity of threads matters for collaboration. Not only inactive threads, but an insufficient number of threads to constitute complete warps, raises corner cases in which original computation should be retained. Our shuffle synthesis handles both situations by adding dynamic checkers for uniformity.

Listing~\ref{lst:synthesis} presents an example of the synthesis by
PTXASW. Once all the emulation is finished, the results are collected
and filtered to satisfy all the above-mentioned conditions. Then,
PTXASW selects the possible shuffle for each load with the smallest
shuffle delta ($N$) and allows only the least corner cases.  %
At the code generation, each source load instruction is extended to be accompanied by the \texttt{mov} instruction to prepare the source register (\texttt{\bfseries\color{emphpink}\%source}).
The destination load is covered with the shuffle operation and a corner-case checker.
First, we check if the thread has no source from the same warp (\texttt{\%out\_of\_range}). Second, the incompleteness of the warp (\texttt{\%incomplete}) is confirmed with a warp-level querying instruction.
In any case, the shuffle operation is performed at the position of the original load, shifting the value of the source register with the distance of the extracted shuffle delta.
Finally, only the threads participating in an incomplete warp or assuming no source lane execute the original load under the predicate (\texttt{\%pred}).
When $N<0$, the \texttt{shfl} instruction takes the \texttt{.up} option and when $N>0$, the \texttt{.down} option is selected.
If $N=0$, just the \texttt{mov} instruction is inserted instead of all the synthesized code.
In actual code, the calculation of \texttt{\%warp\_id} is shared among shuffles and set at the beginning of the execution to reduce the computational latency.

To preserve the original program characteristics, such as the register
use, uniformity, and ILP, following ways of generation are avoided.
We can produce the correct results even if \texttt{shfl} is predicated by \texttt{\%incomplete}, but it often imperils the basic efficiency with an additional branch, which limits ILP.
On the other hand, our code introduces only one predicate to each shuffle and does not leave any new branch in the resultant SASS code.
Also, we do not use a select instruction for merging the results
between shuffles and corner cases, because it would aggravate register pressure.
The output predicate by shuffle poses execution dependency and provides the invalid status of inactive threads; thus, it is ignored.
Moreover, we only create shuffles from direct global\-/memory loads and do not implement shuffles over shuffled elements for better ILP.

\vspace{-0.2cm}
\section{Experimental Methodology}\label{sec:method}

We build PTXASW using Rosette~\cite{rosette}, a symbolic\-/evaluation system upon the Racket language.
PTXASW is equipped with a PTX parser and runs the emulation of the parsed code while expressing runtime parameters as symbolic bitvectors provided by Rosette.
Our shuffle synthesis is caused at code generation, which prints the assembler-readable code.
We evaluate our shuffle mechanism with the NVHPC compiler~\cite{nvhpc} by hooking the assembler invocation and overwriting the PTX code before it is assembled.
The NVHPC compiler accepts the directive-based programming models
OpenACC and OpenMP to generate GPU code, which have no control over warp-level instructions.
The emulation is also tested for GCC with OpenACC/OpenMP code and LLVM with OpenMP code, but these use a master-worker model to distribute computation across thread-blocks~\cite{llvm} and do not directly refer to the thread ID in each thread, so mainly ineffective results are obtained.
Our synthesis is not limited to global-memory loads and works on shared memory (such as Halide~\cite{halide}), but the performance is not improved due to the similar latency of shared-memory loads and shuffles.
The NVHPC compiler utilizes the same style to translate both OpenACC
and OpenMP codes written in C/C++/Fortran to PTX, hence supporting any combinations.

\begin{table}[t]
\center
\begin{tabular}{c|ccccc}
name & Lang & Shuffle/Load & Delta & Analysis \\\hline
\textbf{divergence} & C & 1 / 6 & 2.00 & 4.281s \\
\textbf{gameoflife} & C & 6 / 9 & 1.50 & 3.470s \\
\textbf{gaussblur} & C & 20 / 25 & 2.50 & 7.938s \\
\textbf{gradient} & C & 1 / 6 & 2.00 & 4.668s \\
\textbf{jacobi} & F & 6 / 9 & 1.50 & 4.119s \\
\textbf{lapgsrb} & C & 12 / 25 & 1.83 & 14.296s \\
\textbf{laplacian} & C & 2 / 7 & 1.50 & 4.816s \\
\textbf{matmul} & F & 0 / 8 & - & 13.971s \\
\textbf{matvec} & C & 0 / 7 & - & 4.929s \\
\textbf{sincos} & F & 0 / 2 & - & 1m41.424s \\
\textbf{tricubic} & C & 48 / 67 & 2.00 & 1m39.476s \\
\textbf{tricubic2} & C & 48 / 67 & 2.00 & 1m41.855s \\
\textbf{uxx1} & C & 3 / 17 & 2.00 & 7.466s \\
\textbf{vecadd} & C & 0 / 2 & - & 3.281s \\
\textbf{wave13pt} & C & 4 / 14 & 2.50 & 6.967s \\
\textbf{whispering} & C & 6 / 19 & 0.83 & 6.288s \\
\end{tabular}
\vspace{0.1cm}
\caption{\label{tab:bench}The KernelGen benchmark suite. Lang indicates the programming language used (C or Fortran). Shuffle/Load shows the number of shuffles generated among the total number of global-memory loads. Delta is the average shuffle delta. Analysis is the execution time of PTXASW on Intel Core i7-5930K}
\vspace{-1.7em}
\end{table}

For the evaluation, we use the KernelGen benchmark suite for OpenACC~\cite{kernelgen}, shown in Table~\ref{tab:bench}.
Each benchmark applies the operator indicated in the benchmark name, to single or multiple arrays and updates different arrays.
The benchmarks {\bf gameoflife}, {\bf gaussblur}, {\bf jacobi}, {\bf
  matmul}, {\bf matvec} and {\bf whispering} are two-dimensional, whereas others are three-dimensional, both having a parallel loop for each dimension, in which other loops might exist inside---except {\bf matvec}, which features only one parallel loop.
The thread-level parallelism is assigned to the innermost parallel loop and the thread-block level parallelism to the outermost.
We show the total time of running the shuffle\-/synthesized kernel ten times on Kepler (NVIDIA Tesla K40c with Intel i7-5930K CPU), Maxwell (TITAN X with Intel i7-5930K), Pascal (Tesla P100 PCIE with Intel Xeon E5-2640 v3), and Volta (Tesla V100 SXM2 with IBM POWER9 8335-GTH). We use NVHPC compiler 22.3 with CUDA 11.6 at compilation, but due to environmental restrictions, run the programs using CUDA driver 11.4/11.4/10.0/10.2 for Kepler/Maxwell/Pascal/Volta, respectively.
The compiler options in NVHPC are \texttt{"-O3 -acc \seqsplit{-ta=nvidia:cc(35|50|60|70),cuda11.6,loadcache:L1}"}.
To fully utilize computation, 2D benchmarks select 32768x32768 as
their dynamic problem sizes and 3D compute 512x1024x1024 grids, except {\bf
  uxx1}, which leverages 512x512x1024 datasets and {\bf whispering}, where more buffers are allocated,
computing over 8192x16384 data elements.
To assess a performance breakdown, we prepare two other versions of
PTXASW: NO LOAD and NO CORNER. The former eliminates loads that are
covered by shuffles, whereas the latter only executes shuffles instead of original loads, without the support of corner cases.

The shuffle synthesis fails on four benchmarks.
In {\bf matmul} and {\bf matvec}, the innermost sequential loop contains loads, but these do not have neighboring accesses along the dimension of the thread ID.
The benchmarks {\bf sincos} and {\bf vecadd} do not have several loads sharing the same input array.

\vspace{-0.2cm}
\section{Evaluation}\label{sec:evaluation}

\begin{figure}[t]

  \centering
  \includegraphics[width=0.48\textwidth]{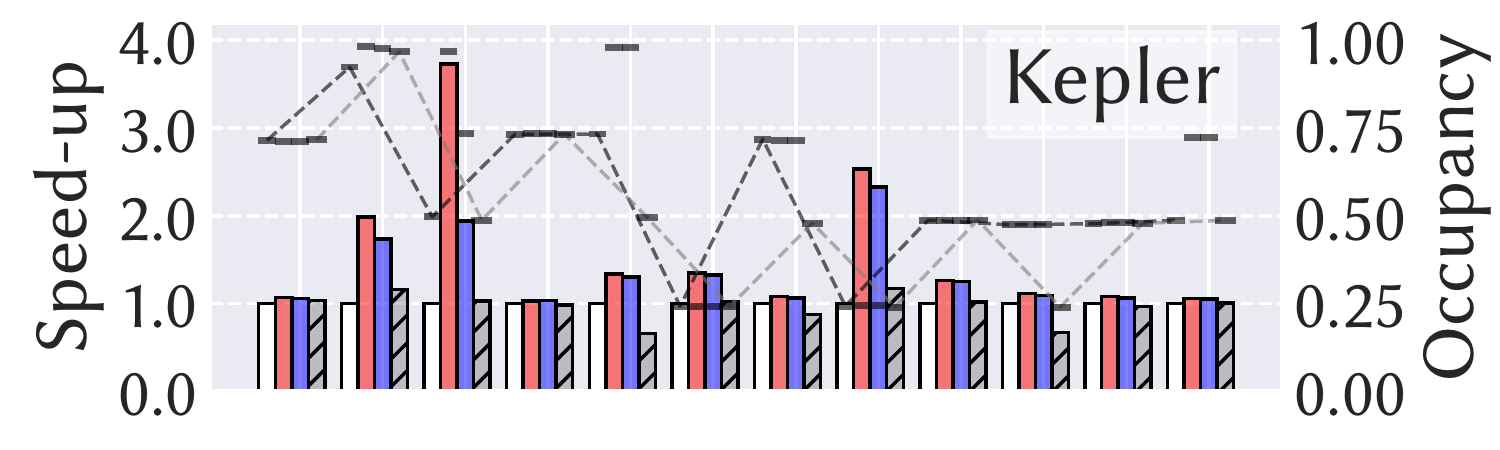}

  \vspace*{-0.15cm}

  \includegraphics[width=0.48\textwidth]{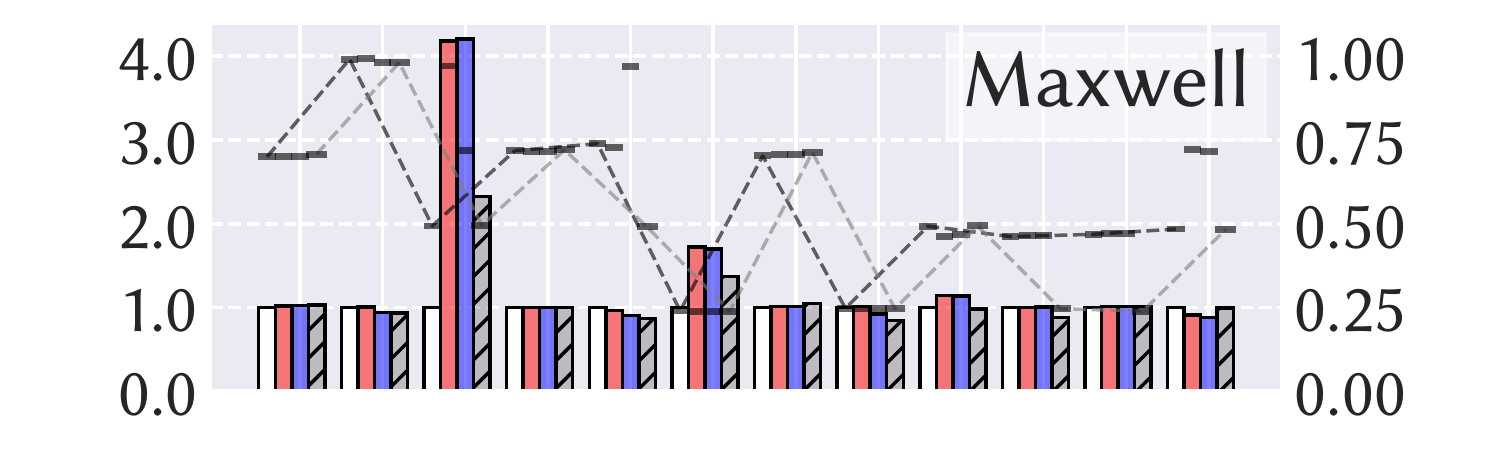}

  \vspace*{-0.15cm}

  \includegraphics[width=0.48\textwidth]{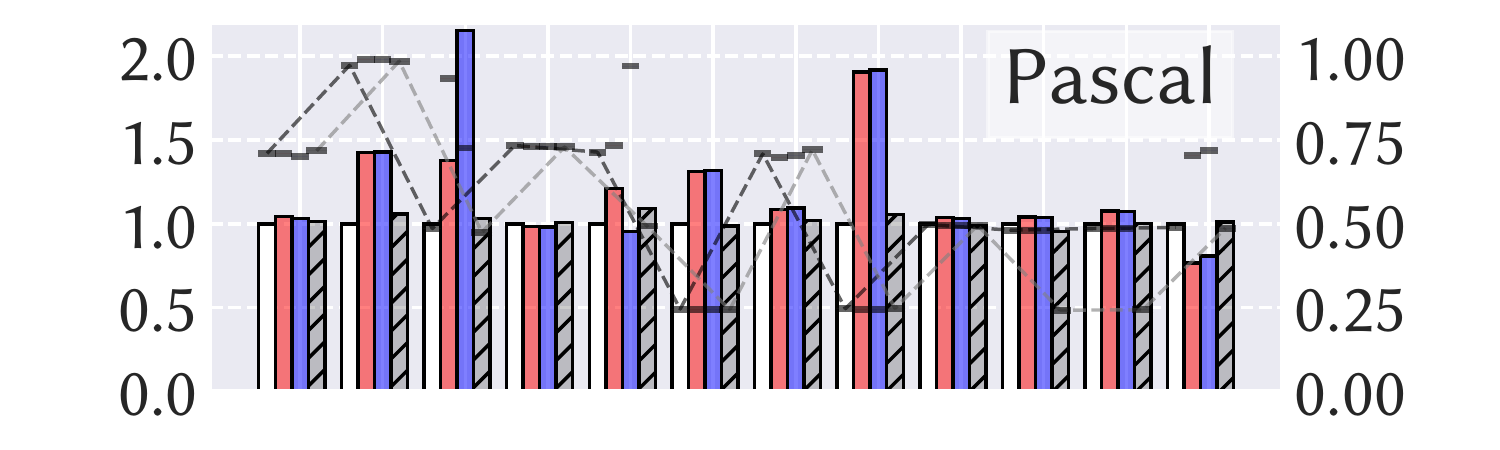}

  \vspace*{-0.15cm}

  \includegraphics[width=0.48\textwidth]{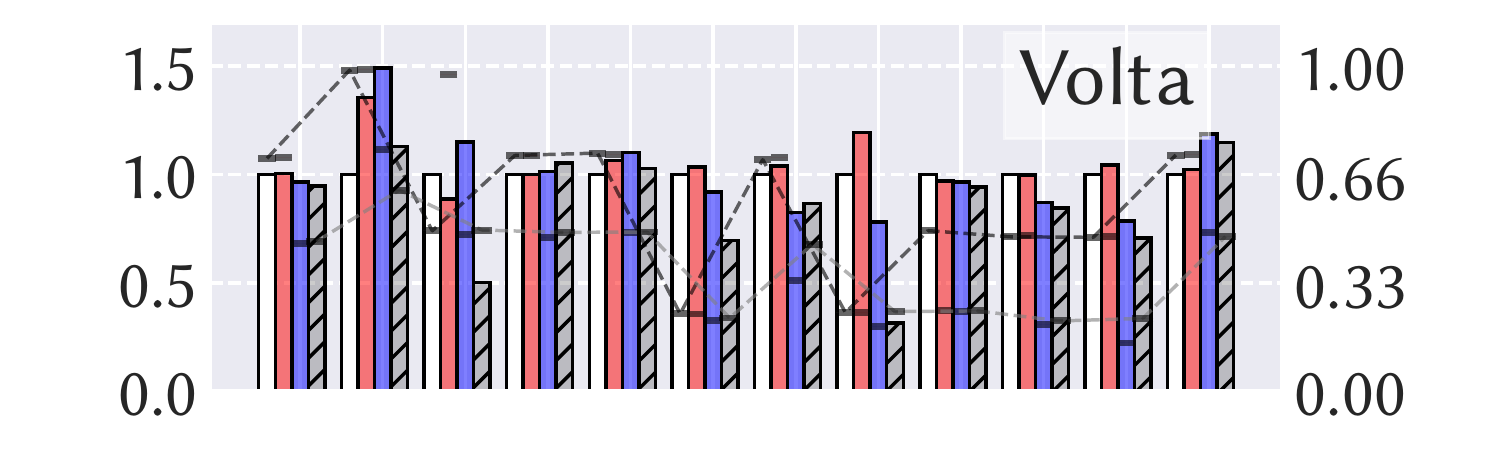}

  \vspace*{-0.225cm}

  \includegraphics[width=0.5\textwidth]{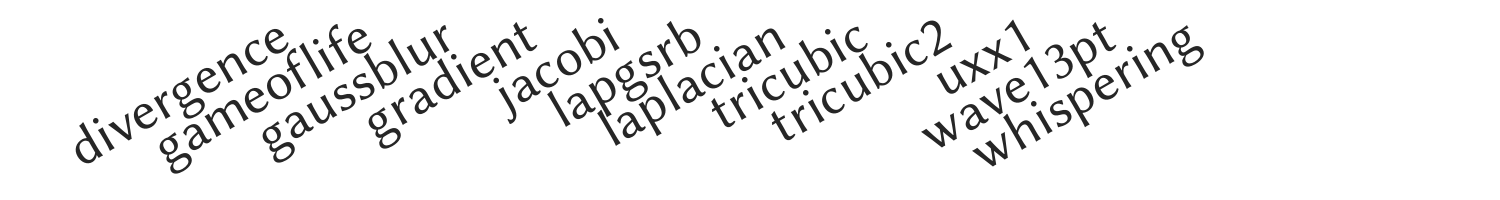}

  \vspace*{-0.1cm}

  \includegraphics[width=0.46\textwidth]{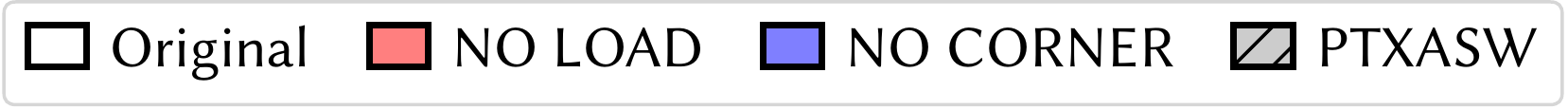}

  \vspace*{-0.2cm}

  \caption{Speed-up compared to Original. NO LOAD/NO CORNER produce invalid results}
  \label{fig:speedup}
  \vspace*{-0.1cm}
\end{figure}

Figure~\ref{fig:speedup} shows the speed-ups of benchmarks on each GPU with original code and PTXASW-generated code along with the NO LOAD and NO CORNER versions. 
The line plots provide the SM occupancy of each benchmark.
Since there is no resource change other than the register use from the
original execution, the occupancy rate is directly affected by the number of registers.
The performance improvement on Kepler/Maxwell/Pascal/Volta is confirmed with 7/6/9/4 benchmarks showing up to 16.9\%/132.3\%/9.1\%/14.7\% performance improvement, respectively.
We see performance degradation with Volta in the case where more than ten shuffles are generated.
Other GPUs mostly gain better performance with such cases.
With increased shuffle deltas, more corner cases are expected.
Volta shows optimal efficiency when $N \leqslant 1.5$, while other GPUs benefit from the case of $N = 2.5$. For example, Maxwell attains the best performance with {\bf gaussblur} ($N = 2.5$), although Volta's performance drops by half for the same case.
The average improvement across all GPU generations is -3.3\%\myslash
10.9\%\myslash 1.8\%\myslash -15.2\% for Kepler\myslash Maxwell\myslash
Pascal\myslash Volta, respectively.

Overall, the performance improvement by PTXASW is found when NO LOAD and NO CORNER have sufficiently better performance compared to the original and when the occupancy typically rises on Kepler/Maxwell and drops on Pascal/Volta.
The average number of additional registers with NO LOAD/NO CORNER/PTXASW compared to the original is -6.4/-5.2/2.7 on Kepler, -6.6/-5.9/4.2 on Maxwell, -7.0/-5.9/3.8 on Pascal, and -6.4/6.8/9.2 on Volta.

\vspace{-0.2cm}
\section{Analysis}\label{sec:analysis}

This section provides the performance detail of our shuffle synthesis on each GPU. Figure~\ref{fig:stall} shows the ratio of stall reasons sampled by the profiler for all the benchmarks.
Those characteristics of computation appear as the results of the program modification (e.g. register use, shuffle delta) and the architecture difference (e.g. computational efficiency, cache latency).

\vspace*{-0.1cm}
\subsection{Kepler}

The Kepler GPU has long stalls on computational operations with each
benchmark. The average execution dependency is 24.7\% and pipeline
busyness is 7.5\% with the original. %
When we look at the memory-bound benchmarks such as {\bf gameoflife},
{\bf gaussblur}, and {\bf tricubic}, NO LOAD significantly reduces the
amounts of memory-related stalls. Especially, {\bf tricubic} has 56.0 percentage points below memory throttles from the original to NO LOAD, yielding 2.53x performance.
From NO LOAD to NO CORNER, the execution dependency increases by 4.0 percentage points and the pipeline busyness decreases by 1.6 percentage points on average. The
performance degradation at NO CORNER with the memory-bound benchmarks
is observed with the latency of the pipelines and the wait for the SM scheduler.
PTXASW suffers from memory throttling and additional computation for the corner cases, which limit the improvement up to 16.9\%.

The memory throttling and the additional computation bottlenecks suffered by PTXASW may be hidden if the shuffle operations reduce the original computation and communication into just one transfer among threads, functioning as a warp-level cache. Otherwise, there is a need to face a trade-off between the redundancy of operations and the efficiency on the architecture.
On Kepler, both heavy computation and memory requests are imposed by the corner case.
Therefore, in the general use of shuffles, the uniformity of calculation is crucial and it requires domain\-/specific knowledge.

\subsection{Maxwell}

\begin{figure}[t]
  \centering
  \includegraphics[width=0.48\textwidth]{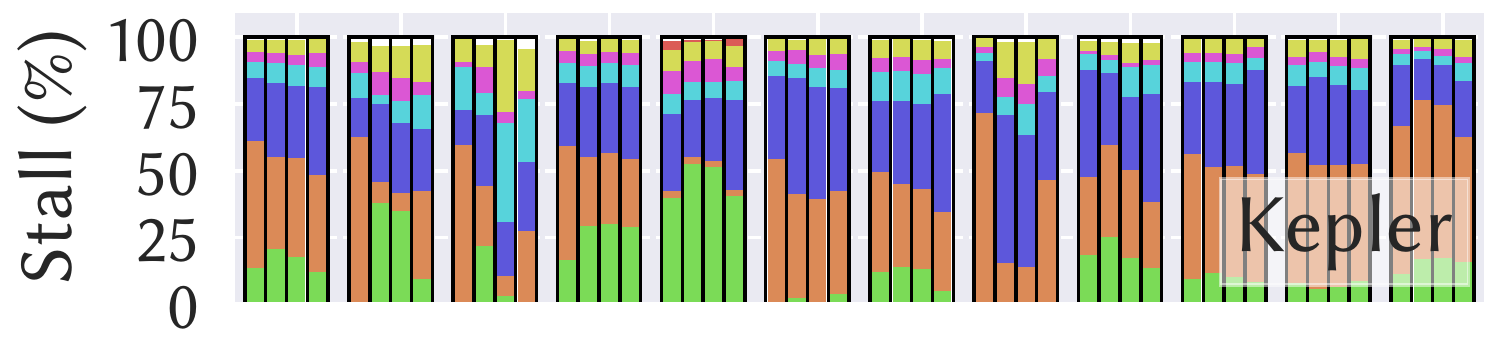}

  \vspace*{-0.15cm}

  \includegraphics[width=0.48\textwidth]{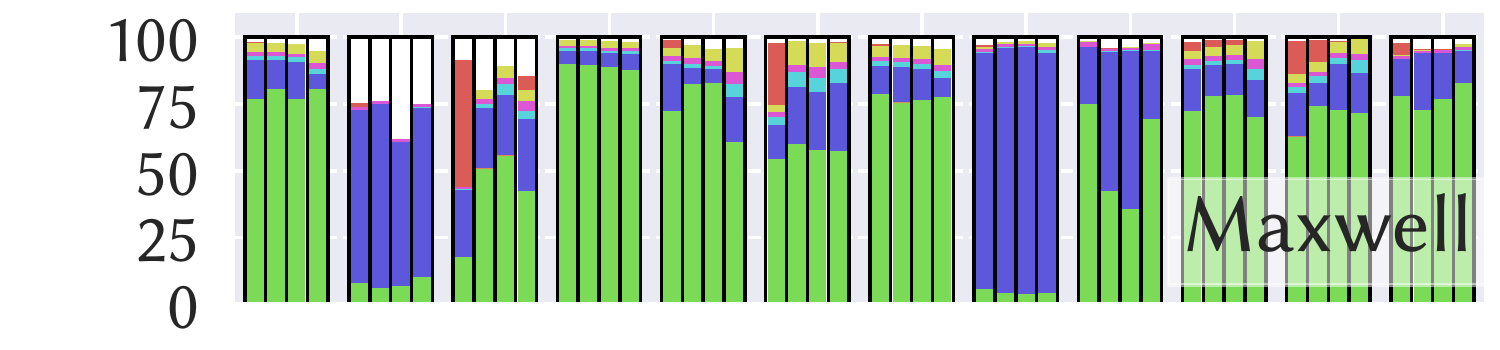}

  \vspace*{-0.15cm}

  \includegraphics[width=0.48\textwidth]{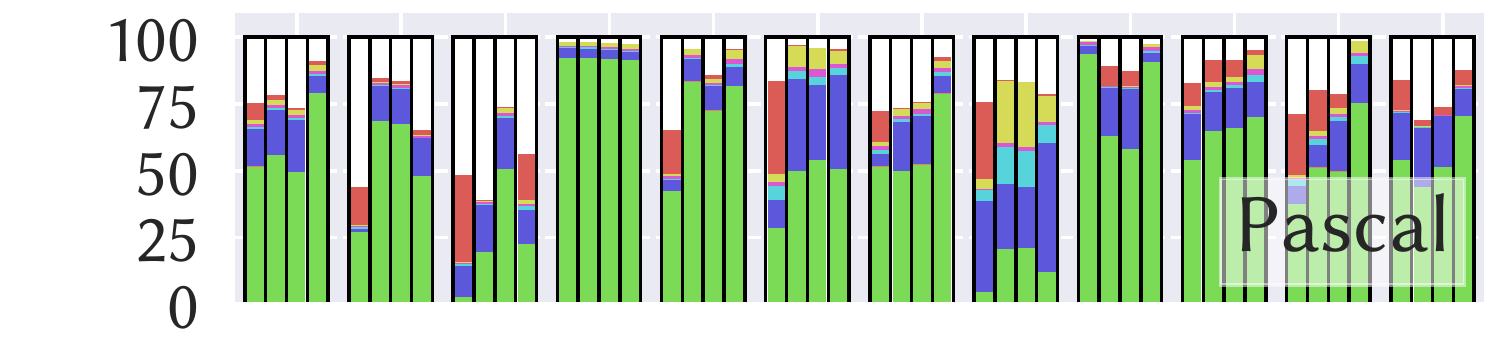}

  \vspace*{-0.15cm}

  \includegraphics[width=0.48\textwidth]{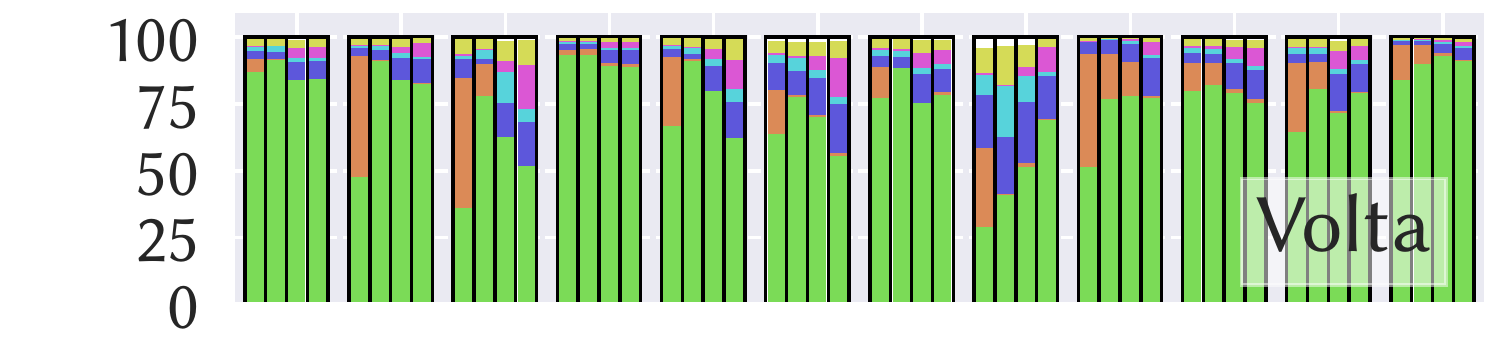}

  \vspace*{-0.15cm}

  \includegraphics[width=0.48\textwidth]{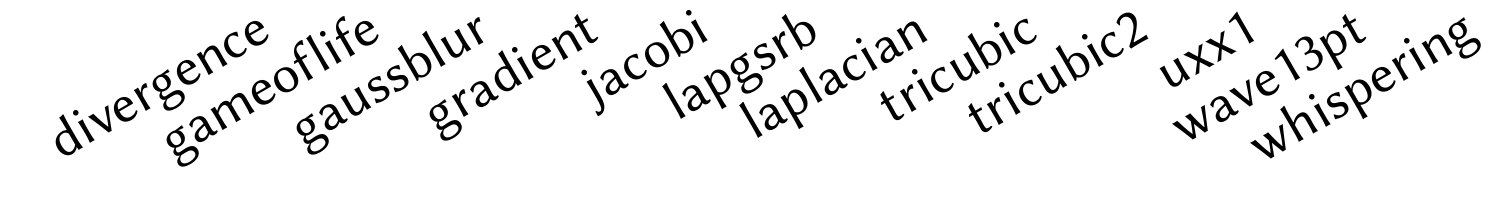}

  \vspace*{-0.1cm}

  \includegraphics[width=0.48\textwidth]{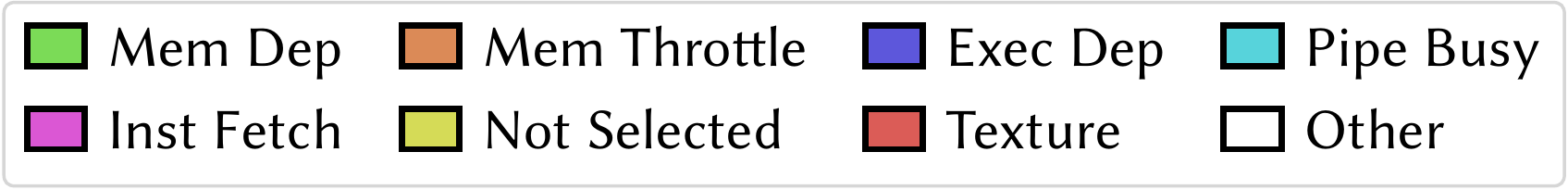}

  \vspace*{-0.2cm}

  \caption{Stall breakdown in the order of Original\myslash NO LOAD\myslash NO
    CORNER\myslash PTXASW from left to right for each benchmark%
  }
  \label{fig:stall}
  \vspace*{-5pt}
\end{figure}

There are two obvious compute-bound benchmarks: {\bf gameoflife} and {\bf tricubic}. For these, no improvement is perceived with NO LOAD, and there are no particular changes in occupancy or stalls throughout the four different versions.
In summary, {\bf gameoflife} experiences -0.1\%\myslash 5.7\%\myslash 6.2\% lower performance and {\bf tricubic} shows -1.6\%\myslash 7.7\%\myslash 15.4\% lower performance with NO LOAD\myslash NO CORNER\myslash PTXASW, respectively, compared to the original version.
In other cases, memory dependency is dominant. However, the merit of
NO LOAD is limited to {\bf gaussblur} and {\bf lapgsrb}, which
experience large texture-memory latency of read-only cache loads,
successfully replaced with shuffles by PTXASW. The texture stall was
reduced from 47.5\% to 5.3\% in {\bf gaussblur} and from 23.0\% to 0.1\% in {\bf lapgsrb} from the original to PTXASW, attaining 132.2\% and 36.9\% higher throughput.
Other benchmarks do not feature stalls that allow for clear
performance improvement by NO LOAD. As it can be observed in
Figure~\ref{fig:stall}, the memory dependency stalls are maintained
for most benchmarks, except for that of {\bf tricubic2}, which shows 32.9 percentage points lower memory dependency and only 14.3\% overall improvement with NO LOAD. Those values are mostly absorbed by the corner cases.

On the Maxwell GPU, only the texture stalls are improvable for efficiency in the tested cases.
Since we observe a moderate overhead of the corner cases, our synthesis tool may enhance the overall performance.
The memory-dependency stalls work as a good indicator of the memory
utilization. If, in addition, a high execution dependency would exist, it would provide the warp-level shuffle optimization the opportunity to be beneficial to speed up the computation.

\vspace*{-0.1cm}
\subsection{Pascal}

Even more than in Maxwell, texture stalls are found in most benchmarks and those produce higher throughput with NO LOAD. Especially, {\bf gameoflife} and {\bf tricubic}, the compute-bound kernels on Maxwell, become memory intensive on Pascal and the performance increases by 5.9\% and 5.4\% with PTXASW.
The unspecific latency ("Other") fills many parts of computation on Pascal.
Further investigation shows that this mainly consists of the latency
from register bank conflicts and the instructions after
branching. With the optimization adding a predicate to check the
activeness of the warp (\texttt{@!incomplete}) before the shuffle and
generating a uniform branch, the ratio of this latency improves from 34.4\% to 8.6\% with PTXASW at {\bf gameoflife}, obtaining 150.8\% efficiency compared to the original.
However, as mentioned in Section~\ref{subsec:codegen}, it decreases
the average relative execution time to 0.88x slowdown.

Since the latency of the L1 cache is higher than that of one shuffle operation, the computation may be hidden by data transfers.
Once the memory\-/dependency stall ratio increases due to replacing the texture stalls, Pascal may maintain the efficiency with the corner cases, resulting in speed-up in nine benchmarks.
For shuffle instructions to be beneficial, the execution should be less divergent and careful register allocation is recommended to maximize the thread utilization.

\subsection{Volta}

On Volta, most benchmarks become memory-bound and memory-intensive
applications become sensitive to memory throttles. Nevertheless, the
speed-up by NO LOAD is limited to up to 1.35x ({\bf gameoflife}), due to the highly efficient cache mechanism. As argued in Section~\ref{sec:evaluation}, some of the benchmarks attain higher performance with NO CORNER than in the case of NO LOAD for the lower occupancy.
Other than that, we observe performance degradation due to increased execution dependency for {\bf lapgsrb} and {\bf tricubic} with NO CORNER.
Those further reduce the efficiency with PTXASW while featuring stalls for instruction fetching.
Also, the memory dependency of {\bf tricubic} develops a large latency for memory accesses with PTXASW even though the corner cases experience fewer loads.
This leads to unstable speed-ups between 0.315x and 1.15x.

The calculation through shuffles is expected to be effective depending on the utilization of communication, and the nonentity of warp divergence.
Especially, as Volta shows minimal latency at each operation,
the penalty of non-aligned computation becomes apparent and must be avoided by the algorithm.

\vspace*{-0.1cm}
\subsection{Application Example}

We also apply PTXASW for the compilation of CUDA benchmarks extracted from applications. We select three benchmarks that appeared as complex 3D stencil operations in~\cite{rawatgpu}: {\bf hypterm}, {\bf rhs4th3fort}, and {\bf derivative}, to run on the Pascal GPU.
{\bf hypterm} is a routine from a compressible Navier-Stokes mini-app~\cite{exact}.
{\bf rhs4th3fort} and {\bf derivative} are stencils from geodynamics seismic wave SW4 application code~\cite{sw4}.
Each thread in the benchmarks accesses 152\slash 179\slash 166 elements over 13\slash 7\slash 10 arrays, respectively.
We modify the execution parameters to execute at least 32 threads along the leading thread-block dimension and use the float data type.
Since we saw in the prior section the overhead of long-distance shuffles, which generate many corner cases, we limited the shuffle synthesis to be $|N| \leqslant 1$ and found shuffles only with $|N| = 1$.

{\bf hypterms} contains three kernels that work along different dimensions. In the kernel for the leading dimension, 12 shuffles are generated over 48 loads,
producing 0.48\% improvement.
{\bf rhs4th3fort} and {\bf derivative} feature a single kernel each.
{\bf rhs4th3fort} experiences 2.49\% higher throughput by PTXASW while placing 44 shuffles %
among 179 loads.
For {\bf derivative}, having 52 shuffles %
from 166 loads, PTXASW attains 3.79\% speed-up compared to the original execution.

\vspace{-0.2cm}
\section{Related Work}\label{sec:related}

Ever since warp-shuffle instructions were introduced during the Kepler generation of GPUs, these have been the subject of various lines of research. Early work described their manual use for specific computational patterns such as reduction operations~\cite{reduction} and matrix transposition~\cite{tra}. Other research described the use of warp-shuffle instructions in the context of domain-specific optimizations such as employing them as a register cache for stencil operations~\cite{register-cache}, or to replace memory access for Finite Binary Field applications~\cite{register-cache}.

Research on the automatic generation of warp-shuffle instructions has been explored. Swizzle Inventor~\cite{swizzle} helps programmers implement swizzle optimizations that map a high-level "program sketch" to low-level resources such as shuffle operations. The authors meticulously design the abstraction of shuffles, synthesize actual code roughly based on algorithms found in previous literature, and attain enhanced performance while reducing the amounts of computation.
Tangram, a high-level kernel synthesis framework, has also shown the
ability to automatically generate warp-level
primitives~\cite{simon2019}. Unlike the work presented in this paper,
both of the above-mentioned efforts leverage domain-specific
information to map computational patterns such as stencil, matrix
transposition, and reductions to shuffle operations.

Recent code-generation techniques allow for obtaining optimal SIMD
code generation. Cowan {\it et al.}~\cite{cowan2020} generate program
sketches for execution on ARM processors, by synthesizing additional
instructions, as well as input/output registers, to implement the
shortest possible SIMD code of reduction. Unlike PTXASW, which uses an SMT solver to find the optimal shuffle deltas, this work runs a comprehensive search of multiple possible code versions; thus, the search space is exponential to the number of instructions.
VanHattum et al.~\cite{dsp} attain faster execution on digital signal processors while employing {\it equality saturation}~\cite{equality}, a modern way of optimization that generates possible code as much as possible from a basic program according to the rules of term rewriting.
They derive shuffles along with vector I/O and computation from sequential C code.
Their intermediate code contains instructions in one nested expression and the shuffle operation only works for memory loads that appear as arguments of the same vector operation. Therefore, the code rewriting for shuffles assumes a top-down style where outer expressions have to be vectorized first, in order to vectorize inner expressions containing shuffled loads.
While their technique may provide a powerful method to the implementation of libraries, irregular patterns such as corner cases
found in HPC applications are out of scope.

\vspace{-0.1cm}
\section{Conclusion}\label{sec:conclusion}

This paper introduces symbolic emulation to compiling GPU code in order to discover hidden opportunities for optimization.
We employ several languages, enabling OpenACC directives such as in C and Fortran, for the frontend to generate GPU assembly code. Then, our tool emulates the code upon symbols that substitute dynamic information. While pruning control flows to reduce the emulation time, we automatically find possible warp-level shuffles that may be synthesized to assembly code to bypass global-memory accesses. We apply this technique to a benchmark suite and complex application code showing results that improve multiple benchmarks on several generations of GPUs. We also provide the latency analysis across multiple GPUs to identify the use case of shuffles.

\vspace{-0.1cm}
\section*{Acknowledgement}
We are funded by the EPEEC project from the European Union’s Horizon 2020 research and innovation program under grant agreement No. 801051 and the Ministerio de Ciencia e Innovación—Agencia Estatal de Investigación (PID2019-107255GB-C21/AEI/10.13039/501100011033). This work has been partially carried out on the ACME cluster owned by CIEMAT and funded by the Spanish Ministry of Economy and Competitiveness project CODEC-OSE (RTI2018-096006-B-I00).

\newpage
\balance
\bibliographystyle{ACM-Reference-Format}
\bibliography{main}

\end{document}